\input harvmac


\def\half{{1\over 2}}

\def\kbT{k_{\scriptscriptstyle\rm B}T}

\def\bold#1{\setbox0=\hbox{$#1$}%
     \kern-.010em\copy0\kern-\wd0
     \kern.025em\copy0\kern-\wd0
     \kern-.020em\raise.0200em\box0 }

\lref\BERNE{C.\ Pangali, M.\ Rao, and B.\ J.\ Berne, Chem. Phys.
Lett.\ {\bf 55}, 413 (1978).}
\lref\DF{M.~Dijkstra and D.~Frenkel, Phys. Rev. E {\bf 51}, 5891 (1995).}
\lref\SS{F.~T.~Wall and F.~Mandel, J. Chem. Phys. {\bf 63}, 4592 (1975);
F.~Mandel, J. Chem. Phys. {\bf 70}, 3984 (1979).}
\lref\KAS{J. K\"as, unpublished.}
\lref\SB{J.~V.\ Selinger and R.~F.\ Bruinsma, Phys. Rev. A. {\bf 43}, 2910
(1991).}
\lref\KLN{P. Le~Doussal and D.~R.\ Nelson, Europhys. Lett. {\bf 15}, 161
(1991);
R.~D.\ Kamien, P. Le~Doussal, and D.~R.\ Nelson, Phys. Rev. A {\bf
45}, 8727 (1992); Phys. Rev. E {\bf 48}, 4116 (1993).}
\lref\AWM{X.~Ao, X.~Wen and R.~B.\ Meyer, Physica A {\bf 176},
63 (1991).}
\lref\MEYER{R.B.~Meyer, {\sl Polymer Liquid Crystals}, edited
by A.~Ciferri, W.~R.\ Kringbaum and R.~B.\ Meyer (Academic, New York, 1982)
Chapter 6.}
\lref\DG{P.G.~de~Gennes, {\sl Polymer Liquid Crystals}, edited
by A.~Ciferri, W.~R.\ Kringbaum and R.~B.\ Meyer (Academic, New York, 1982)
Chapter 5.}
\lref\TARMEY{V.~G.\ Taratura and R.~B.\ Meyer, Liquid Crystals {\bf 2},
373 (1987).}

\nfig\fone{Single polymer persistence length (in units of the
center to center spacing $a=1.25\sigma$)
as a function of the stiffness
parameter $\kappa$. The solid line shows the persistence lengths measured
from a $1.2\times 10^8$ step run, while the dashed line shows the
persistence lengths measured in the second half of a $2.4\times 10^8$ step
run.}
\nfig\ftwo{Density plots of the polymer structure function
in the $(q_\perp,q_z)$
plane. Here $q_0=4\pi/\sqrt{3}a_0$, where
$a_0$ is the average inter polymer spacing.
Note that there is little scattering along $q_\perp=0$, in agreement
with theoretical arguments.  (a) areal number density $0.142$ (b)
areal number density
$0.32$.}
\nfig\fthree{Two-dimensional in-plane structure function for areal number
densities $0.142$ and $0.32$.  This is obtained by $S_2(q_\perp) =
\int {dq_z\over 2\pi}
S(q_\perp,q_z)$.  The curves are successively shifted along the
intensity axis for clarity.}

\Title{IASSNS-HEP-95/94}{\vbox{\centerline{
Structure Function of Polymer Nematics:}\vskip2pt\centerline{A Monte
Carlo Simulation}}}

\centerline{Randall D. Kamien}
\smallskip\centerline{\sl School of Natural Sciences,
Institute for Advanced Study, Princeton, NJ 08540 \rm and}
\centerline{\sl Department of
Physics and Astronomy, University of Pennsylvania, Philadelphia, PA 19104}
\smallskip\centerline{and}\smallskip\centerline{Gary S. Grest}\smallskip
\centerline{\sl Corporate Research Science Laboratory}
\centerline{\sl Exxon Research and Engineering, Annandale, NJ 08801}
\vskip .3in
We present a
Monte Carlo simulation of a polymer
nematic for varying volume fractions, concentrating on the structure function
of the sample.  We achieve nematic ordering with stiff polymers made
of spherical monomers that would otherwise not form a nematic state.
Our results
are in good qualitative agreement
with theoretical and experimental predictions, most notably the bowtie pattern
in the static structure function.  \hfill\break
\Date{21 December 1995; revised 22 September 1996}

There has been considerable interest recently in a variety of liquids
composed of line-like objects aligned on average along a common direction.
Flux lines in high-temperature superconductors, strings of electric
dipoles in electrorheological fluids and polymer nematics are all
systems with this common morphology.  Despite the vast literature
on nematic liquid crystals, the number of theoretical and experimental
studies of polymer nematics
is relatively small.  Building on work by Meyer \MEYER\ and de Gennes \DG ,
statistical mechanical treatments of polymer nematics \refs{\SB,\KLN}\
have predicted, in particular, the structure function
in the semi-dilute regime.  Ao, Wen and Meyer \AWM\ have
studied the structure of
poly-$\gamma$-benzyl-glutamate in the nematic phase by X-ray
scattering. However because the experiments cannot reach very small momentum
$q$, due to the forward scattering beam stop, they cannot
test many of the main
theoretical predictions which are good at these
small $q$ values.

We have simulated polymer nematics with the hope of
gaining some additional information in the behavior of the
polymer nematics.  In our simulation
nematic order arises {\sl only} due to the polymer stiffness -- there is
no microscopic, steric nematic interaction.
We are particularly interested
in the small $q$ region of the static structure function.
The resulting equilibrium polymer configurations could be
used to compare to actual polymer configurations in nematic solutions
of actin \KAS.   We contrast this study to recent work \DF\ which studied
the phase behavior of $10$-mers made of rigid nematogens.  In our simulation
nematic order arises {\sl only} due to the polymer stiffness -- there is
no microscopic, steric nematic interaction.
Because of the long length scales and slow relaxation times
which are inherent in any polymer problem, it is simply not
feasible to model in great detail the chemical complexity of
any of the experimental polymer nematics such as actin.
Fortunately since we are interested in understanding some
of the basic properties of polymer nematics, particularly
at small $q$ (large distances), it is not necessary to include
the chemical detail explicitly. For this reason, we chose to study
a coarse grained pearl-necklace model similar to that previously
studied for flexible polymer systems \ref\BAUM{A.~Baumg\"artner, Ann. Rev.
Phys. Chem. {\bf 35} 419 (1984); {\sl Applications of the Monte Carlo
Method in Statistical Physics}, edited by K.~Binder (Springer, Berlin,
1984).}.
The stiffness of the chain was controlled by adding
a three-body bending term which could be tuned to vary
the persistence length of our polymer nematic relative to
its chain length. Using this model, we are able to test
several of the theoretical predictions of \refs{\SB,\KLN}.

We studied an ensemble of $M$ polymer nematics in a periodic
rectangular cell. Each monomer of the polymer was modeled
by a hard sphere of diameter $\sigma$ connected to its neighbors
along the chain with a fixed bond of length $a=1.25\sigma$.
Results presented here were for $M=200$ chains of length $N=50$.
Since we are only interested in the static equilibrium properties
of the system and not their dynamics, we used a slithering
snake algorithm \SS\ which is very efficient particularly when
the persistence length is large \ref\footn{We attempted to study this
system via a molecular dynamics simulation.  We found that equilibration took
an
unacceptably long time.}. The algorithm models the reptation
motion of a snake by attempting to remove the head (tail) of the chain
and placing it at the tail (head) with a random bond angle.
We modeled the bond energy as $-\kappa \cos \theta$ where $\theta$
is the bond angle, and $\kappa$ is the bending stiffness measured in
units of $\kbT$.
If the move violates the excluded volume constraint on the beads,
the move is rejected. If not, then the Monte Carlo move is
accepted according to the relative energies of the original
and new bond angle. The results presented here are from runs made using
a force-biased algorithm \BERNE\ in which moves are only rejected due
to excluded volume.  While more moves were accepted compared to
a standard Monte Carlo algorithm, the additional
complexity did not lead to significant time reduction at high density
on either the
Silicon Graphics Power Challenge
or the IBM Power PC604 (runs were done on both). 
Since we were interested in rather stiff
chains and the rejection rate for the moves was quite high, we
quote all times in terms of successful moves \ref\footn{
On the Power PC, a 
force-biased run of 120 million successful steps 
took roughly 1000 hours 
at
areal number
density $\rho_0\sigma^2={\hbox{\sl number of polymers}\over\hbox{\sl area in
units of $\sigma^2$}}=0.32$
(areal fraction, $\rho_0\pi\sigma^2/4=0.25$;
volume fraction $\phi=0.13$).  On a 143MHz Sun UltraSPARC a force-biased run at
$\rho_0\sigma^2=0.142$ took roughly 1000 hours.
On the
Power Challenge (with 75MHz IP21 processors) a non force-biased run of 120
million successful moves took 700 hours for areal number density
$\rho_0\sigma^2=0.569$
(areal fraction, $\rho_0\pi\sigma^2/4=0.45$;
volume fraction $\phi=0.25$). For this
case, the acceptance rate is only about $0.3\%$. This low rate was
due to both the rejection rate caused by the bond angle and also the
high-densities leading to a small free volume. }.
Because of the long runs we concentrated on
three areal number densities $\rho_0\sigma^2=0.142$
(areal fraction, $\rho_0\pi\sigma^2/4=0.11$), $0.32$
($0.25)$, and $0.569$ $(0.45)$.  At the highest density the system did
not equilibrate and presumably became glassy.  We will not report here on the
highest density run.
We found that the autocorrelation of the end-to-end distance for
a single polymer at the bending stiffness we are interested in ($\kappa=40$)
decayed exponentially with a relaxation time of about $1.5\times10^5$ Monte Carlo
moves.  Thus we believe that after $1.2\times 10^8$ steps the two lower density
runs
should have equilibrated. The simulations were started from an initial
condition
in which all $200$ polymers chains were aligned along
the $z$-axis but randomly placed
in the $xy$ plane at random heights along the $z$-axis.
At $\rho_0\sigma^2=0.142$,
each monomer, on average, moved $15\sigma$ {\sl laterally}, while at
$\rho_0\sigma^2=0.32$
each moved $10\sigma$.
The chains were placed in a periodic rectangular box with
a fixed height $L_z=Na=62.5\sigma$ in the $z$-direction and
a variable $xy$ area.   For the densities we ran at, we found a Maier-Saupe
order
parameter ($S=[3\langle\cos^2\theta\rangle-1]/2$) of $0.049$ and $0.849$
for $\rho_0\sigma^2 = 0.142$ and $0.32$, respectively, where $\theta$
is the angle between the polymer bond and the principal axis with the
largest eigenvalue of the bond-moment
tensor $T_{ij} = \langle\hat t_i\hat t_j\rangle$, where $\hat t$ is the vector
pointing along a bond.
Since there was no explicit nematic field
added, there is nothing to stop the ordered domain from rotating to a new
direction.  In the denser run we found that the nematic axis differed from
the $\hat z$-axis by approximately $7^{\circ}$.  In the most dilute system,
since
there was no nematic order, the issue was moot.  We thus
take the $z$-direction to be the ordering direction for the
analysis.  In addition, since the starting
states had $S=1$, the lower average values of the order parameter give us an
independent
confirmation that the system has equilibrated.

We calibrated the bending energy $\kappa$ with polymer persistence length
for a dilute chain. The  persistence length was determined
from the bond angle-bond angle correlation length.
In Fig.~1, we show the
persistence length $L_p/a$ as a function of $\kappa$ calculated
from  the first half (solid
line) and second half (dashed line) of a $2.4\times 10^8$ step run.
Because of the excluded volume interaction, a monomer cannot bend
back on its neighbor and the maximum bending angle of a single
bond is $133^{\circ}$.  This is why the persistence length $L_p/a$ is
greater than $1$ for $\kappa=0$.  In addition, note that if we were to take the
continuum limit, the bending energy would become $\half\kappa\int(\partial_s\theta)^2$.  
This simple theory leads to a persistence length $L_P/a=\kappa/\kbT$, in agreement
with our calibration when the stiffness is such that self-avoidance is not an issue.
In an attempt to match to the experiments of Ao, Wen and Meyer \AWM\ where
the polymer persistence length was $80\%$ of the chain length and
what can be run in a reasonable amount of time, we concentrate
on $\kappa=40$ for the rest of this paper. This
corresponds to a persistence length $L_P/a\approx 39$ and $L_p/[(N-1)a]=80\%$.

Let us recall the theoretical expectations for the structure
function \refs{\SB,\KLN}
\eqn\estrere{S(q_\perp,q_z) = {\langle\,\delta\rho(q_\perp,q_z)
\delta\rho(-q_\perp,-q_z)\,\rangle\over \rho_0^2} ,}
where $q_\perp$ is the magnitude
of $\vec q_\perp\equiv(q_x,q_y,0)$,
$\rho=\rho_0+\delta\rho$ is the local areal number density and
$\rho_0$ is the average mean density.
For infinitely long polymers, we expect
no scattering along the $q_z$ axis due to the conservation of
polymer density along the $z$-axis.  As described by Taratura and
Meyer \TARMEY , the projection of the polymer tangent into the
$xy$-plane $\vec t$ and the areal density $\rho$ satisfy
\eqn\econst{\partial_z\rho + \nabla_{\!\perp}\!\cdot\!\rho\vec t =0}
When $q_\perp=0$ \econst\ implies that $\partial_z\rho(q_\perp=0,z)
=0$.  Hence there is no density contrast and there will
be no scattering along the $q_\perp=0$ axis. When the polymers are
finite in extent, \econst\ is modified by adding sources and sinks
for polymer heads and tails to the right hand side of the equation.
This will lead to some scattering along the $q_z$ axis \KLN\ controlled by
the typical polymer length.
In addition we expect Bragg-like peaks on the $q_\perp$ axis, corresponding
to the incipient columnar crystal being formed by the polymers.  Together
these two predictions suggest the typical bowtie anisotropic
pattern of X-ray scattering.
The average static scattering function $S({\bf q})$  can easily be determined
for the simulated system from
\eqn\sq{S({\bf q})={1\over NM}\big<|\sum_i\exp(i{\bf q\cdot r_i})|^2\big>.}
We took the average every $1.2\times 10^6$ moves.
Because of the periodic boundary conditions each $q_i$ has to be commensurate
with the dimensions of the cell. This means that the smallest nonzero
$q_i=2\pi/L_i$, where $i$ is one of the Cartesian coordinates.
Thus as $\rho_0$ increases at fixed $M$, the smallest accessible $q_i$
increases, which is one reason we did not study higher density systems.
In Fig.~2, we present the results for the two dimensional structure
factor $S(q_{\perp},q_z)$ for the two densities studied.

Our data agree qualitatively with the first expectation and quantitatively with
the second.
In addition we calculated the average polymer persistence length for the
densities considered.  We find $L_P/a = 39$ ($\rho_0\sigma^2=0.142$) and $95$
($0.32$).  This is consistent with the increased
amount of nematic ordering
with increasing density.  Note that the value at the lowest density is close
to the dilute value of $39$.

While the structure function data in the $(q_\perp,q_z)$ plane does not
have enough resolution for the fitting of the
full two dimensional surface, some information
may be gleaned from collapsing the data onto the $q_\perp$ axis.  Indeed,
the derived function
\eqn\esii{S_2(q_\perp) = \int {dq_z\over 2\pi}\,S(q_\perp,q_z)= \int
{dq_z\over 2\pi}\,e^{iq_zz}S(q_\perp,q_z)\bigg|_{z=0} = S(q_\perp,z=0)}
is the structure function of the polymer nematic in any fixed $z$
cross section \KLN.
Since the numerically determined structure function is
only computed for $q_z=-100\pi/L_z\ldots 100\pi/L_z$ (where $L_z=Na=62.5\sigma$
is
the height of the box) the sum performed to calculate $S_2(q_\perp)$
will be somewhat smaller than the actual value of $S_2(q_\perp)$.
We see the formation of true liquid-like structure in the nematically ordered
system and of almost gas-like order in the isotropic system.  This
is certainly reasonable: the isotropic system is in no sense a directed
line-liquid and we should not expect any correlated behavior in a constant
$z$-slice.
Note that in the ordered run $S_2(q_\perp)$ has
maxima at multiples of
$q_0 = 4\pi/(\sqrt{3}a_0)$ where $a_0$ is the average inter-polymer
spacing, indicative of the incipient crystalline order.

In this study we have presented the first Monte Carlo study of
a nematic polymer liquid crystal for densities in the semi-dilute
regime. We found that for intermediate densities, the static
structure function is anisotropic and has the predicted bowtie shape
in the $(q_{\perp},q_z)$ plane.
Using a sum-rule, similar to that relating the structure function at zero
momentum to the bulk compressibility \KLN\ we could
extend this work
to study the scaling of the bulk modulus over a range of densities.
In addition, the sum rule could be checked differently by varying
the average polymer length.  Polydispersity should not effect the
theoretical predictions: this could be checked as well.
Finally, a study which varied the persistence length at fixed density
would also be enlightening and could be compared to theory.

It is a pleasure to acknowledge stimulating discussions with P.~Heiney, J.~K\"as,
F.~MacKintosh,
R.B.~Meyer and D.R.~Nelson.  The help of T.~Tysinger is especially 
appreciated.  RDK would like to thank Exxon Research
and Engineering where some of this work was done.
RDK was supported in part by the National Science Foundation, through Grant
No.~PHY92--45317 (IAS), through Grant No.~DMR91-22645
and through the National Scalable Cluster Project,
Grant No. CDA94-13948
(University of Pennsylvania).

\listrefs
\listfigs

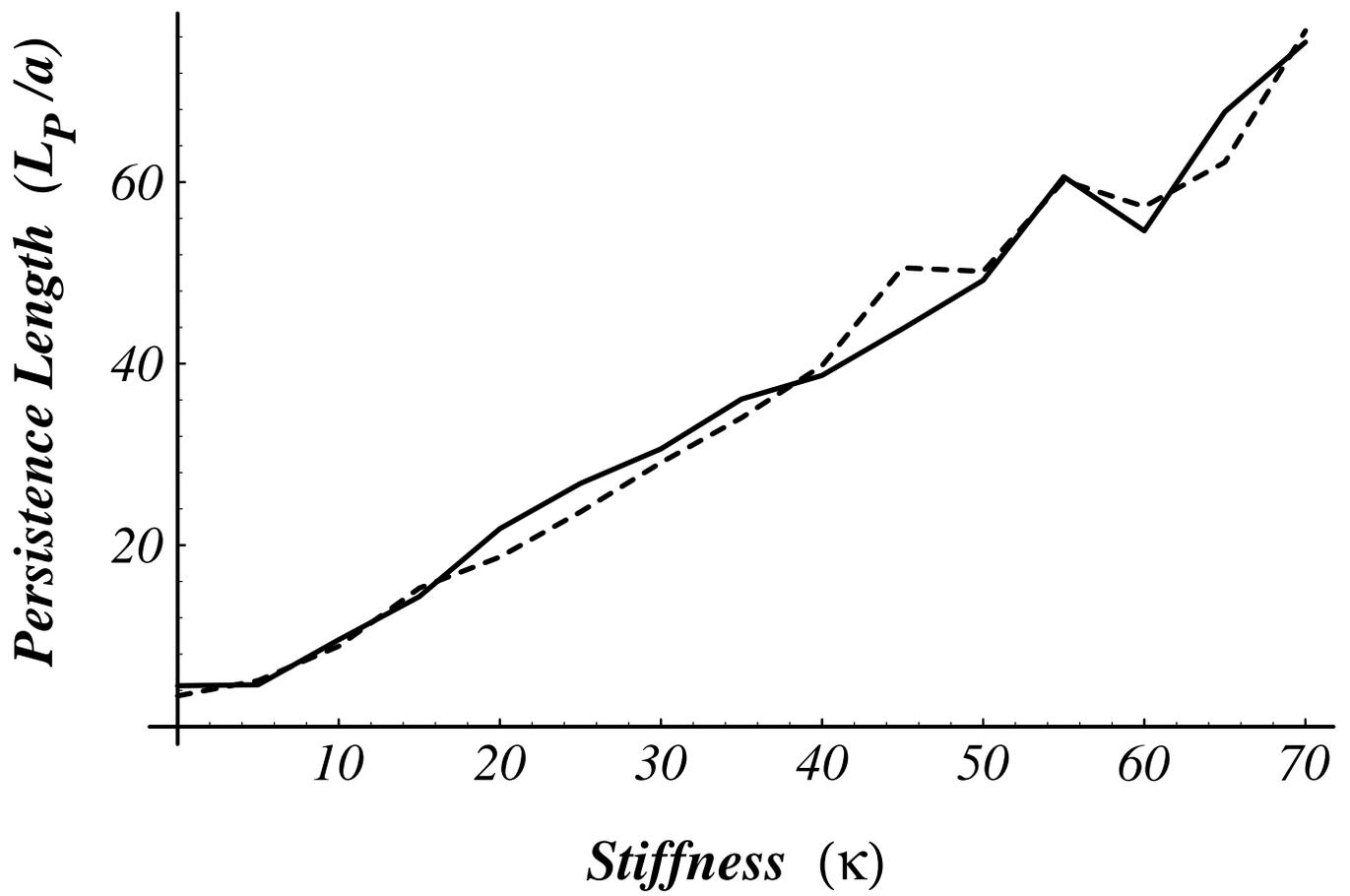
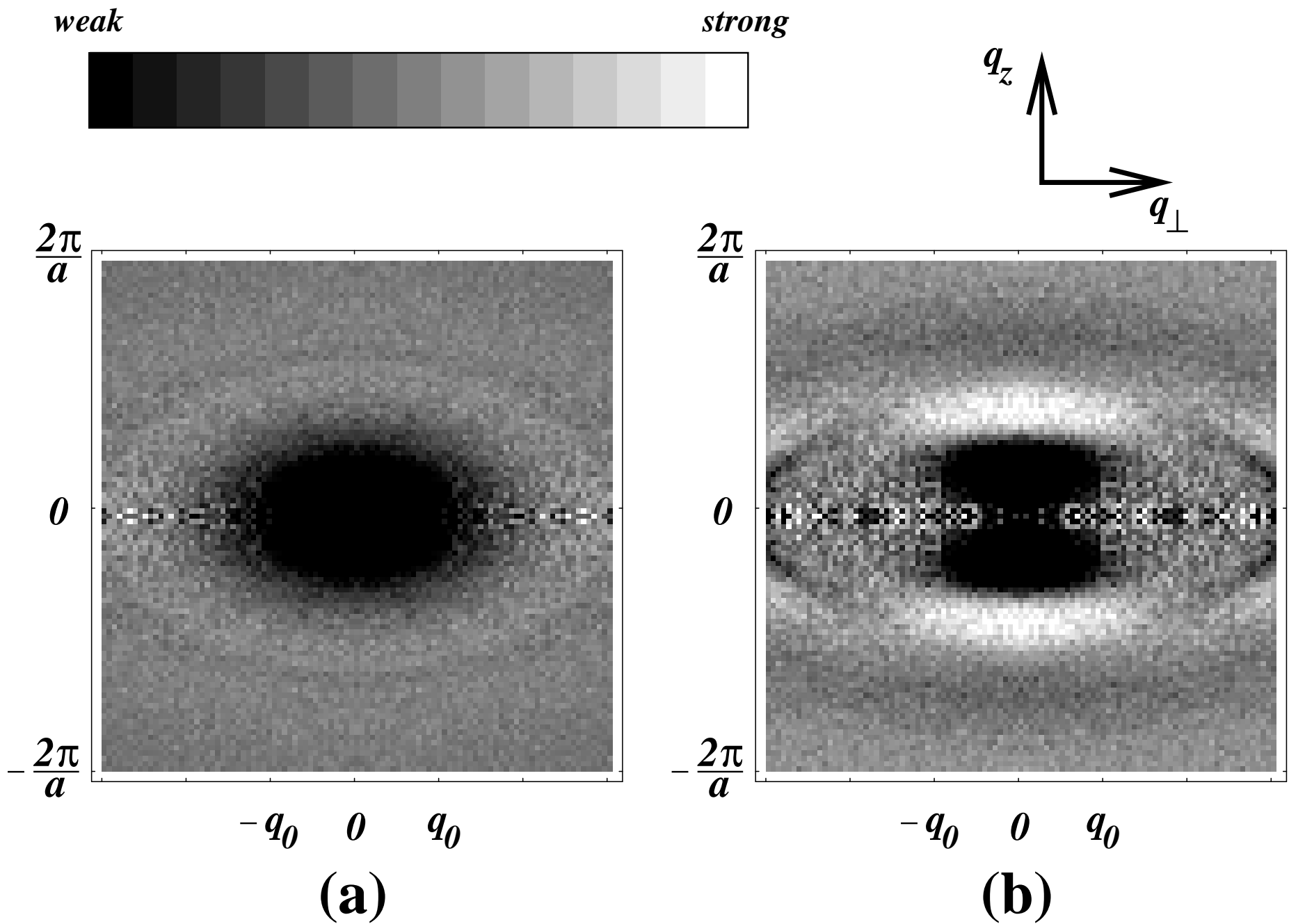
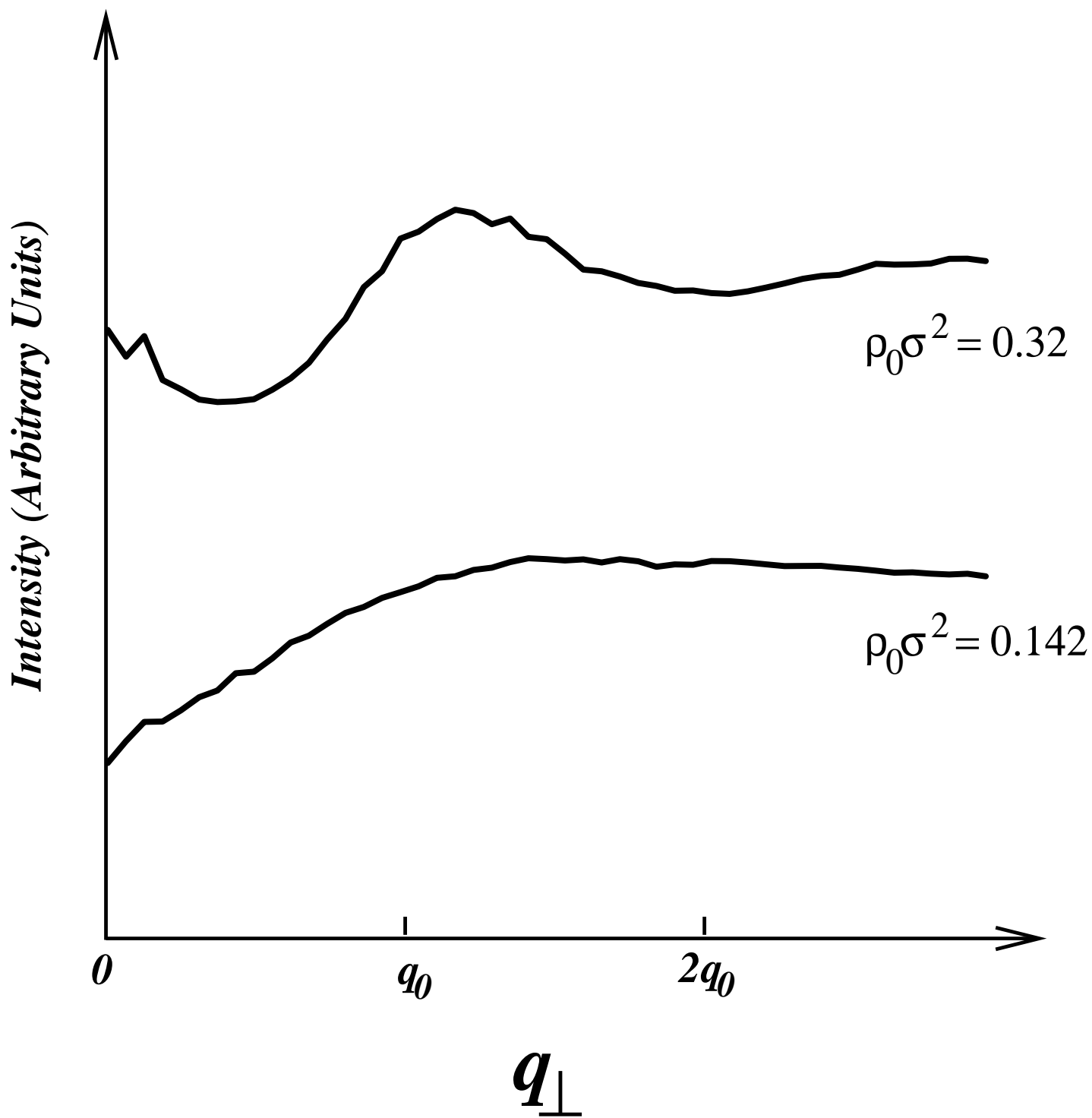
\bye